# The Effects of Pinning on Critical Currents of Superconducting Films

R. P. Simões, P. A. Venegas and D. F. Mello

Faculdade de Ciências, Unesp-Universidade Estadual Paulista, Departamento de Física, Bauru, SP, CP 473, 17033-360, Brazil

**Abstract** – Using molecular dynamics simulations, we analyze the effects of artificial periodic arrays of pinning sites on the critical current of superconducting thin films as a function of vortex density. We analyze two types of periodic pinning arrays: hexagonal and Kagomé. For the Kagome pinning network we calculate, using two directions of transport current: longitudinal and transversal. Our results show that the hexagonal pinning array presents higher critical currents than Kagomé and random pinning configuration for all vortex densities. In addition, the Kagomé network shows anisotropy in their transport properties.

## 1. INTRODUCTION

Considerable efforts have been directed toward improving critical current density ($J_c$) in superconductors through the development and application of various manufacturing techniques. Pinning of vortices is one of the most commonly employed mechanisms used to increase and maintain the critical current density, which will depend on the efficacy of the pinning mechanism. As a result, many attempts have been made to enhance the pinning of vortices in high temperature superconductors (HTSCs) by creating structural defects using various sample treatments, especially that of energetic radiations. By using lithography it is possible to create well-defined periodic nanostructured arrays with well-defined periodic pinning structures in which the microscopic pinning parameters, such as size, depth, periodicity, and density, can be carefully controlled. This enhancement of critical currents using periodic arrays has recently been demonstrated for high $T_c$ superconductors [1-11]. Therefore, the study of pinning mechanisms and vortex motion, from both the experimental and theoretical point of view, is important to understand and to create materials with more applicability.

In the present work we analyze the vortex motion and the behavior of the critical current when the density of vortices is changed. The calculations are made for a thin superconducting film with hexagonal and Kagomé periodic pinning networks. Kagomé periodic pinning arrays present important and interesting characteristics [3]. They have been extensively used to study several physical phenomena and give us the possibility to study some of their unique features [3]. Furthermore, experimental results [11] suggest that when there are interstitial vortices, in superconductors with periodic pinning arrays, the transport properties in two perpendicular directions can be anisotropic. In the present work we analyze the effects of the anisotropy in the critical currents for the Kagomé pinning lattice using two mutually perpendicular directions of transport current and compare with the critical current of the hexagonal lattice. In the following study, we present the model and numerical simulation for the pinning lattices under consideration and analyze how the critical current behaves for a wide range of vortex densities.

## 2. MODEL

In order to study the dynamical properties of the vortex lattice we use the molecular dynamic technique and the system of vortices is described by a Langevin equation at zero temperature. The equation of motion for a vortex at the position $\mathbf{r_i}$ is written as:

$$\frac{d\mathbf{r_i}}{dt} = \sum_{i \neq j} \mathbf{f_i^v}(r_{ij}) + \sum_p \mathbf{f_i^p}(r_{ip}) + \mathbf{F} \quad (1)$$

Using the notation of Ref. [4], $\mathbf{f_i^v}(r_{ij}) = -C_v \nabla \ln(r_{ij})$ is the repulsive force between vortex $i$ and vortex $j$ at a distance $r_{ij}$, and $\mathbf{f_i^p}(r_{ij}) = -A_p \nabla \exp\text{-}(r_{ip}^2)$ the attractive force between vortex $i$ and the pinning center $p$ at a distance $r_{ip}$. The third term, $\mathbf{F} = \phi_0/c(\mathbf{J} \times \mathbf{z})$, is the driving force associated to the transport current $\mathbf{J}$. $A_v$ and $A_p$ are the vortex-vortex and vortex-pinning interaction strengths respectively. The length scales are normalized by $4\xi$, the energy scales by $A_v$, and the time by $16\eta\xi^2/A_v$. In the present calculation our system corresponds to a thin film which is infinite in the transverse ($x$) and longitudinal ($y$) directions (see Fig. 1). The driving force $F$ is applied in longitudinal direction for the hexagonal lattice and longitudinal and transversal direction for the Kagomé lattice. In Fig. 1 we show the Kagomé lattice where the rate between the number of the vortices ($N_v$) and the number of pinning centers ($N_p$) is $N_v/N_p = 4/3$.

For the simulation we used a system with 144 vortices where the vortex density $n_v = N_v/L_xL_y$ in a box of size $L_x$ by $L_y$ was varied from 0.01 to 0.5 vortices per unit length, which is the range where the critical current shows an appreciable change. The density is increased in steps of 0.1, and for each step we start the simulation process by relaxing the vortex lattice during 2000 time steps without the transport current, in order to find the most stable configuration. This configuration was used as our initial boundary condition. Then, the transport current is applied, and the time average of the vortex velocity and differential resistance are calculated over 30000 time steps, in order to obtain the critical force (or current) values. For characterizing dynamical phases of the vortex flow, we also analyze the vortex trajectories, structure factor and vortex diffusion of each vortex density.

pinning lattice and longitudinal force. Two different phases are characterized: I) Below the critical force we have the pinned phase, II) Above the critical force, there is a sharp transition to a moving crystal, all vortices start to move simultaneously at the same velocity and there is no diffusion between the different vortex channels. An analog result, not shown here, is obtained for the hexagonal pinning lattice with transverse transport current but their critical force is lower than the longitudinal case.

In figure 2b we can see the result for Kagomé lattice with longitudinal applied force. The vortex system shows three phases: I) Below the critical force vortices are in the pinned phase, II) Above the critical force half of the vortices start to move in longitudinal channels, while the other part remains trapped, III) By increasing the force, the other half of the vortices begin to move but with smaller velocity and with no diffusion between channels.

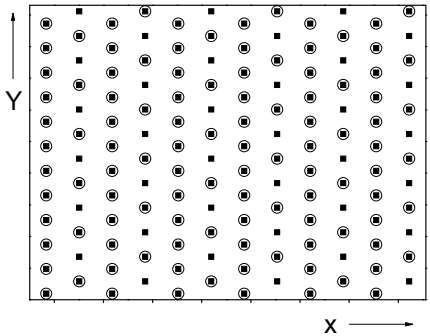

**Figure 1.** In the Figure the black dots represent the hexagonal vortex lattice and the open circles the Kagomé pinning network.

## III. RESULTS AND DISCUSSION

The critical forces can be obtained from the analysis of the time average of vortex velocity as a function of the applied transport force. On Figures 2a, b and c we show the vortex velocities for the hexagonal and Kagomé lattices and a selected vortex density $n_v=0.12$. The changes in the vortex velocity suggest different dynamic phases for each pinning array, however, the complete characterization of each phase is made by means of the differential resistance, vortex trajectories and structure factor. Note that the analysis of our results for other vortex densities shows analogous phase transitions but the forces that define the boundary between the different phases are different. Figure 2a shows the results for hexagonal

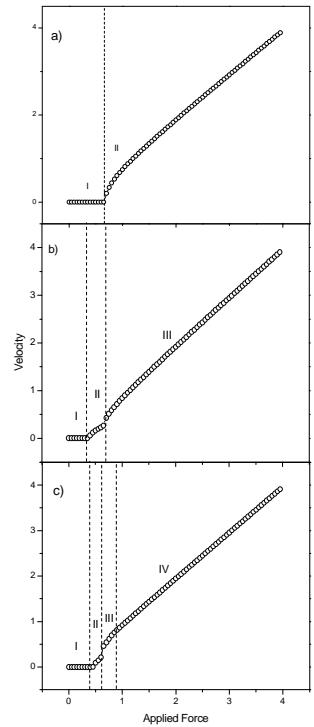

**Figure 2.** The Figure sows the time average of vortex velocity for a) Hexagonal lattice, b) Kagomé lattice with transport force on longitudinal direction and c) Kagomé lattice with transport force on transversal direction.

Figure 2c shows the results for Kagomé Lattice and transversal transport force. In this case, four different phases can be characterized: I) Below the critical force the vortices are in the pinned phase; II) Above the critical force the vortices present a complex dynamic behavior, vortices jump occurs to alternate rows, i.e. a given vortex jumps to the right row and the other to the left row, forming a sinuous trajectory; III) Increasing the force, all the vortices are moving in almost straight channels with interconnectivity. The analysis of the structure factor, shows that this dynamic regime resembles the smectic flow observed in simulation with random pinning arrays of vortices described in Ref. [4]; IV) In this phase the driving current is so strong and the vortices begin to move in straight channels with no diffusion between them.

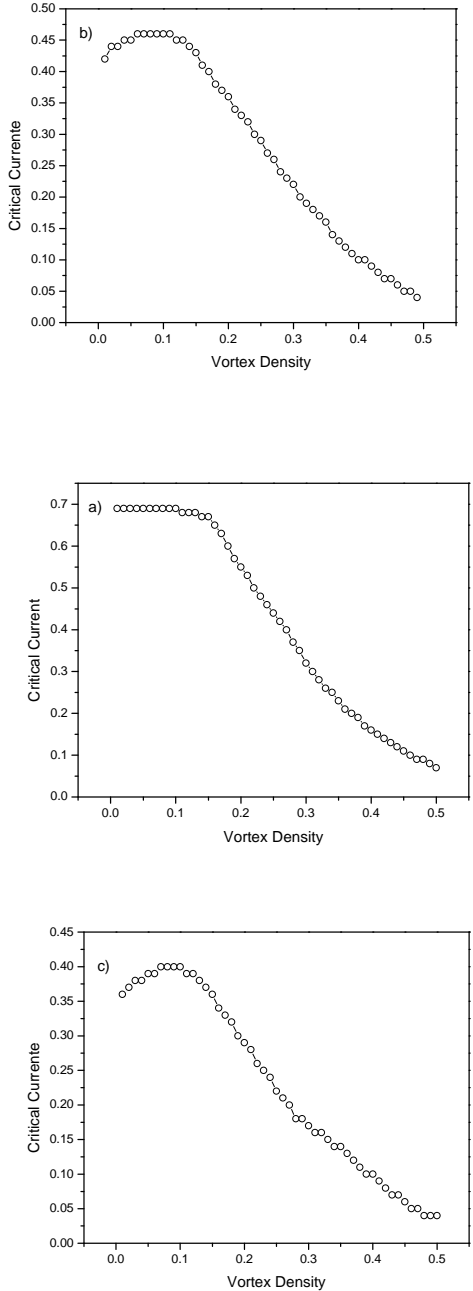

Fig. 4. The figure shows the critical force as a function of the vortex density for: a) a hexagonal pinning network, b) Kagomé pinning network with transversal transport force and c) Kagomé pinning network with longitudinal transport force.

Figures 3a, b and c, show the critical currents for hexagonal and Kagomé pinning networks with longitudinal and transversal transport forces. As it is well stated, our results show that periodic pinning lattices are clearly more effective than random pinning [4] to obtain higher critical currents. From Fig. 3b and 3c we can see that Kagomé Lattice under transversal force presents a higher critical current than longitudinal force, for all vortex density values. That is due to the fact that interstitial pinning is different in each case. However, hexagonal lattice is more efficient for pinning vortices than both cases of Kagomé lattice. This may be understood due to the fact that the Kagomé lattice has a rate $N_v/N_p = 4/3$ and the hexagonal lattice a rate $N_v/N_p = 1$, i.e., the Kagomé lattice is a hexagonal lattice with vacancies. It also becomes evident, that the effectiveness of the pinning mechanisms decreases as the density of vortices is increased for all pinning arrays. That may be explained because for higher densities we are closer to the normal state.

## 4. CONCLUSION

In our study we have compared the critical forces of hexagonal and Kagomé pinning lattices for a wide range of vortex densities. Our results show that the critical force for hexagonal lattice is higher than for Kagomé array of pinning sites. We have shown also the anisotropy in the critical current and dynamical phases of Kagomé pinning lattice when we apply forces that are mutually perpendicular. This result is in accordance with experimental results [11], and can be explained due to the different interstitial pinning in each case.